\magnification 1200
\baselineskip = 15pt
\overfullrule=0pt

\font\Frak=eufm10
\def\frak#1{{\hbox{\Frak#1}}}

\font\Bbb=msbm10
\def\BBB#1{\hbox{\Bbb#1}}

\def\g{{\frak g}}
\def\C{\BBB C}
\def\R{\BBB R}
\def\Z{\BBB Z}
\def\N{\BBB N}

\def\Nodd{\BBB N_{\rm odd}}
\def\Nev{\BBB N_{\rm ev}}
\def\Zodd{\BBB Z_{\rm odd}}
\def\Zev{\BBB Z_{\rm ev}}

\def\d{\partial}
\def\Re{{\rm Re}}
\def\Im{{\rm Im}}
\def\Res{{\rm Res}}

\def\q{{\bf q}}
\def\p{{\bf p}}
\def\tq{{\tilde q}}
\def\tp{{\tilde p}}
\def\tbq{{\bf \tilde q}}
\def\tbp{{\bf \tilde p}}
\def\tx{{\tilde x}}
\def\ty{{\tilde y}}
\def\x{{\bf x}}
\def\arctanh{{\rm arctanh}}
\def\u{U}
\def\v{V}

\def\W{{\cal W}}
\def\Cl{{\cal C}_\infty}
\def\WW{{\W\otimes\W}}
\def\sl{{\widehat{sl_2}}}

\def\bar{\overline}
\def\imply{\Rightarrow}

\centerline
{\bf Sine-Gordon equation and representations}
\centerline
{\bf of affine Kac-Moody algebra $\sl$.}

\

\centerline {\bf Yuly Billig}

\

\centerline{ Department of Mathematics and Statistics,
University of New Brunswick,}
\centerline{  Fredericton, Canada E3B 5A3}

\


\

{\bf 0. Introduction}

\

One of the most fascinating applications of the Kac-Moody
theory is the use of affine Lie algebras and their groups
to exhibit hidden symmetries of soliton equations.
In 1981 M.~Sato [19] and Drinfeld-Sokolov [6] 
(see also [7])
discovered fundamental links between soliton equations 
and infinite-dimensional Lie groups. In an important
sequence of papers [2], [3], [4],
Date, Jimbo, Kashiwara and Miwa gave a construction
of the Kadomtsev-Petviashvili and Korteweg-de Vries hierarchies based on the representation 
theory of affine Kac-Moody algebras. 
The affine algebra $a_\infty$ produces the KP hierarchy,
while the KdV hierarchy is linked to $\sl$.

Date, Jimbo, Kashiwara and Miwa used the vertex 
operator realization of the basic highest weight module,
which was discovered earlier by Lepowsky and Wilson [17].
In this realization for $\sl$, 
the space of the basic module is
identified with the polynomial algebra in infinitely
many variables $\x = (x_1, x_3, x_5, \ldots)$.
A completion of this space is interpreted as 
the space of $\tau$-functions.
In this framework, the B\"acklund transformation that
raises the soliton number of an N-soliton $\tau$-function,
is given by the exponential of a vertex operator.
The Casimir operator equation
$$ \Omega (\tau \otimes \tau ) = 0 \eqno{(0.1)} $$
decomposes in a hierarchy of PDEs in Hirota form.

The sine-Gordon equation
$$ u_{xt} = \sin (u)  \eqno{(0.2)} $$
is of great significance to both mathematics and physics.
This equation appeared first in geometry and describes
the surfaces of constant negative curvature in $\R^3$.
The physical importance of the sine-Gordon equation is
related to the fact that it is a soliton equation which
is manifestly Lorentz-invariant.

The sine-Gordon equation played a special role in 
the development
of the soliton theory. The B\"acklund transformation
method, which is a precursor for the Lie theory approach,
was first developed in the geometric setup for this 
equation.
Before the connection between soliton equations and 
infinite-dimensional Lie algebras was discovered
in 1980's, Mandelstam wrote a paper [18] on
quantized sine-Gordon equation, in which he used
operators somewhat similar to the vertex operators.

From the AKNS method [1] it is known that the sine-Gordon
equation is related to  $\sl$, but the
representation-theoretic interpretation of this
equation was missing.

 The main goal of the present paper is to fill this gap.
Our starting point is the following observation.
From the original Hirota's paper [13] we know that
the sine-Gordon equation (0.2) can be written as a system
of equations in Hirota form on a pair of functions
$(\tau_0, \tau_1)$
$$\left\{ \matrix{
D_x D_t 
(\tau_0 \circ \tau_0 - \tau_1 \circ \tau_1 ) = 0 \cr
 D_x D_t
(\tau_0 \circ \tau_1) =  \tau_0 \tau_1 , \cr}
\right. \eqno{(0.3)}$$
where $u$ is related to $(\tau_0, \tau_1)$ by
$u = 4 \arctan \left( {\tau_1 \over \tau_0} \right)$.
To describe the soliton solutions in terms of
$\tau$-functions, we to put $\tau_0$
and  $\tau_1$ together into $\tau = \tau_0 + i\tau_1$.
Then the $N$-soliton solution for the sine-Gordon
is given by (cf. [13]):
$$\tau = \tau_0 + i\tau_1 = $$
$$ = 
1 + \sum_{k=1}^N \sum_{1\leq j_1 < \ldots < j_k\leq N}
\alpha_{j_1} \ldots \alpha_{j_k} i^k 
\prod\limits_{1\leq r < s \leq k}
\left( {{z_{j_r} - z_{j_s}} \over {z_{j_r} + z_{j_s}}} \right)^2
\exp\left( \left(\sum_{s=1}^k z_{j_s}^{-1} \right) t +  
\left(\sum_{s=1}^k z_{j_s} \right) x
\right)  $$
with real-valued $\alpha_j$'s and $z_j$'s.

If we compare this with the expression for the $N$-soliton
solution for the KdV hierarchy [11], [14]
$$ \tau = 
 1 + \sum_{k=1}^N \sum_{1\leq j_1 < \ldots < j_k\leq N}
\alpha_{j_1} \ldots \alpha_{j_k}  
\prod\limits_{1\leq r < s \leq k}
\left( {{z_{j_r} - z_{j_s}} \over {z_{j_r} + z_{j_s}}} \right)^2
\exp\left( \sum_{m\in\Nodd} \left( \sum_{s=1}^k z_{j_s}^m 
\right) x_m \right) , $$
then it becomes clear that time $t$ in the sine-Gordon
equation is nothing but variable $x_{-1}$ missing in
the KdV picture! The ``correct'' set of variables 
should include both positive and negative odd integers:
$\x = (\ldots, x_{-3}, x_{-1}, x_1, x_3, \ldots )$.
We shall use
the symbols $\Zodd, \Zev, \Nodd, \Nev$ to denote
odd/even integers/natural numbers.

Our extended hierarchy will be completely symmetric
in positive and negative directions. This means that
the $\sl$-module that we need to construct should
not be of the highest/lowest weight. 
However, the Casimir operator is a crucial ingredient 
for the construction of the hierarchy of differential 
equations (see (0.1) above).
Formally, this operator is well-defined
only for the highest/lowest weight modules. Nevertheless,
we are still able to use the Casimir operator in our
situation.

In the vertex operator realization of the basic highest
weight module, $\sl$ acts by differential operators
on the algebra of polynomials $\C[x_1, x_3, x_5, \ldots]$.
In this paper we represent $\sl$ by the same 
differential operators, but now acting on the algebra of
differential operators in variables $x_1, x_3, x_5, \ldots$,
by left multiplication. The algebra of differential 
operators is isomorphic to the Weyl algebra $\W$ generated 
by $p_i, q_j$, \ $i,j \in \Nodd$  with relations
$$ p_i p_j = p_j p_i, 
\hbox{\hskip 0.5cm}
q_i q_j = q_j q_i,
\hbox{\hskip 0.5cm}
p_i q_j - q_j p_i = c_i \delta_{ij} \cdot 1, 
\hbox{\hskip 0.5cm}
c_i \in \C \backslash \{ 0 \} .$$

 Though the Weyl algebra is non-commutative, still the
partial differentiations ${\d \over \d p_i}$,
${\d \over \d q_j}$ are well-defined. This allows us to
view the elements of the Weyl algebra as functions 
on a quantized space.

The Weyl algebra has a Poincar\'e-Birkhoff-Witt decomposition
$$\W = \C [q_1, q_3, \ldots] \otimes \C [p_1, p_3, \ldots] .$$
Since both subalgebras $\C [q_1, q_3, \ldots] $ and
$\C [p_1, p_3, \ldots] $ are commutative, there is a linear
bijection between $\W$ and the polynomial algebra
$\C [\ldots, x_{-3}, x_{-1}, x_1, x_3, \ldots]$:
$$\pi: \W \rightarrow \C [x_1, x_3, \ldots] \otimes
\C [x_{-1}, x_{-3}, \ldots] .$$
The image of a differential operator under map $\pi$ is called
the symbol of a differential operator (see e.g., [16]).

The idea of our approach is to write a Kac-Moody group invariant
equation on a Weyl algebra element $\hat\tau$ and then convert it
into a hierarchy of partial differential equations in Hirota form
on the symbol $\tau = \pi(\hat\tau)$. Then we can use the Kac-Moody
group action to generate the soliton solutions from the trivial
solution $\hat\tau = 1$.

The function $\hat\tau = 1$ is not anymore a solution of (0.1),
since we have changed the action to be the left multiplication
and $1$ now represents the identity operator. However, $\hat\tau = 1$
is trivially a solution of 
$$ \Omega ( \hat\tau \otimes \hat\tau ) -  ( \hat\tau \otimes \hat\tau ) 
\Omega = 0 . \eqno{(0.4)}$$
This algebraic equation transforms into a hierarchy of the non-linear
PDEs in the variables $\x = 
( \ldots, x_{-3}, x_{-1}, x_1, x_3, \ldots)$ which contains two KdV
subhierarchies going in positive and negative directions.

The simplest new equation in this double KdV hierarchy is 
the following (cf. (4.34)):
$$u_{yt} = {\d \over \d x} \left( u_{xxy} + u_x u_y + u_x \right)
\eqno{(0.5)}$$
with $x = x_1, y = x_{-1}, t = x_3$. This is a generalization of 
the KdV with two spatial variables, but different from  the Kadomtsev-Petviashvili equation.

The sine-Gordon equation can not possibly occur in the double KdV
hierarchy (0.4) since the system of Hirota equations (0.3) can not
be expressed in terms of $\tau = \tau_0 + i\tau_1$ alone.

Using the boson-fermion correspondence we construct ``skew''
Casimir operators that split $\tau$ into even and odd parts
(Proposition 2.2). It remains a mystery whether the skew Casimir
operators are related to the Dirac operator. We construct
a hierarchy of equations by considering skew analogs of (0.4)
(see (2.7), (2.8)).

As expected, the sine-Gordon equation with respect to the variables
$x_1$ and $x_{-1}$ appears in it, ``linking'' the two KdV hierarchies.

Here is a brief description of the structure of the paper.
In Section 1 we construct a representation of affine Lie
algebra $a_\infty$ on a completion of the Weyl algebra 
$\W$. We also review the boson-fermion correspondence
and the construction of the Casimir operator. In Section 2
we consider the reduction of this representation to
$\sl$, introduce the skew Casimir operators and
study their properties. In Section 3 we show how 
to convert the skew Casimir operator equations into
a generating series of PDEs in Hirota form and
construct $N$-soliton solutions for these equations.
In the final section we derive non-linear partial
differential equations from the Hirota equations. 
We also obtain the Leibnitz formula for the Hirota
formalism and use it for the treatment of systems
of Hirota equations.

\

{\bf Acknowledgements.} This work is supported by the Natural Sciences and Engineering Research Council 
of Canada. I am grateful to Jack Gegenberg for his
suggestion to study this problem.

\

\

{\bf 1. Boson-fermion correspondence and representations
of $a_\infty$}

\

 In the previous works on applications of 
infinite-dimensional Lie algebras to soliton equations
the main idea is to represent Lie algebras by
differential operators. In this setting the Casimir
operator equation becomes an infinite hierarchy of
non-linear PDEs in Hirota form. The fact that the Casimir
operator commutes with the Lie algebra allows one to 
construct soliton solutions of this hierarchy.

The approach that we take here is to use the action not
on the space of functions, but rather on the Weyl algebra,
or, equivalently, on the space of differential operators
themselves.

 We consider the Weyl algebra $\W$, an associative 
algebra with $1$  generated by the elements $p_i, q_i$,
$i \in \N$ with the defining relations
$$ p_i q_j - q_j p_i = - {4\over j} \delta_{ij} 1,
\hbox{\hskip 0.5cm}
p_i p_j = p_j p_i , 
\hbox{\hskip 0.5cm}
q_i q_j = q_j q_i, 
\hbox{\hskip 1cm} 
i,j \in \N . \eqno{(1.1)}$$
The scaling factor $- {4\over j}$ in the first of these 
relations is chosen to put the vertex operators 
considered below, in a more symmetric form.
 Essentially, $\W$ is a factor of the universal 
enveloping algebra of the Heisenberg algebra in which
the central element is identified with $1$.

The Weyl algebra has a natural representation on the space
of polynomials in infinitely many variables (Fock space)
$F = \C[x_1, x_2, \ldots]$ by the operators of 
differentiation and multiplication:
$$ p_j \mapsto -{2\over j}{\d \over \d x_j},
\hbox{\hskip 0.5cm}
q_j \mapsto 2 x_j ,
\hbox{\hskip 0.5cm}
j \in \N .$$ 
Thus the Weyl algebra can be also interpreted as the
algebra of differential operators on $F$.

In spite of the non-commutativity of the Weyl algebra,
the partial derivatives  ${\d \over \d p_j}, 
{\d \over \d q_j}$ are well defined (which follows
from the fact that the relations (1.1) survive the 
partial differentiation). These partial derivatives
are inner derivations of $\W$ ([5], section 4.6):
$$ {\d \over \d p_j} \left( f(\q, \p) \right) =
 {j\over 4} \left[q_j , f(\q, \p) \right], 
\hbox{\hskip 0.5cm}
 {\d \over \d q_j} \left( f(\q, \p) \right) =
 - {j\over 4} \left[p_j , f(\q, \p) \right], 
\hbox{\hskip 0.5cm}
f(\q, \p) \in\W . \eqno{(1.2)}$$
Here $\q = (q_1, q_2, \ldots)$, \ 
$\p = (p_1, p_2, \ldots).$

The exponentials $\exp (\alpha {\d \over \d p_j}) , {\ }
\alpha\in\C$, (resp. $\exp (\alpha {\d \over \d q_j})$)
are the shift operators 
\break
$p_i \mapsto p_i + \delta_{ij} \alpha$,
$q_i \mapsto q_i$, (resp.  $p_i \mapsto p_i$,
$q_i \mapsto q_i + \delta_{ij} \alpha$), which are 
well-defined commuting automorphisms of $\W$.

Clearly, an element $f(\q,\p)\in\W$ can be written in many
ways. We call the normal form of $f(\q,\p)$ its presentation
in which the generators $q_i$'s are grouped to the left
of $p_j$'s in all monomials. The normal form
can be found using the standard Poincar\'e-Birkhoff-Witt
procedure. It also corresponds to the decomposition
 $\W = \W^- \otimes \W^+$, where $\W^-$ (resp.
$\W^+$ ) is an abelian algebra generated by
$q_1, q_2, \ldots$ (resp. $p_1, p_2, \ldots)$.

The subalgebra $\W^-$ (resp. $\W^+$) is isomorphic 
to the algebra of polynomials in infinitely many
variables $\C[x_1, x_2, \ldots]$ 
(resp. $\C[x_{-1}, x_{-2}, \ldots]$ ). Consider the 
isomorphism of vector spaces
$$ \pi: \hbox{\hskip 0.5cm} \W^- \otimes \W^+
\rightarrow \C[x_1, x_2, \ldots
x_{-1}, x_{-2}, \ldots] ,$$
$$ \pi \left( f(q_1, q_2, \ldots) g(p_1, p_2, \ldots) 
\right) = f(x_1, x_2, \ldots) g(x_{-1}, x_{-2}, \ldots).$$
The map $\pi$ coincides with the notion of
a symbol of a differential operator expressed in  
the Weyl algebra language. This map allows us to convert the elements of the
Weyl algebra into ordinary functions in commuting
variables.

We define the normally ordered product \ $:fg:$ \ 
of $f(\q,\p)$ and $g(\q,\p)$ to be
$$ : f(\q,\p) g(\q,\p) : = \pi^{-1} \left(
\pi( f(\q,\p)) \pi( g(\q,\p)) \right) .$$

We will consider a $\Z$-grading of 
$\W = \mathop\oplus\limits_{n\in\Z} \W_n$ 
by assigning degrees to the generators as follows:
$$ \deg (p_i) = i, \hbox{\hskip 0.3cm} \deg (q_i) = -i, 
\hbox{\hskip 0.5cm} 
i\in\N .$$
Next we introduce a completion $\bar\W$ of the Weyl
algebra, in which the vertex operators will be later
defined. We set the completion of $\W_n$ to be
$$ \bar \W_n = \prod_{{i\leq 0, j\geq 0} \atop i+j=n}
\W^-_i \otimes \W^+_j \eqno{(1.3)}$$
and define $\bar\W$ as
$$ \bar\W = \mathop\oplus\limits_{n\in\Z} \bar \W_n .
\eqno{(1.4)}$$
It follows from the Poincar\'e-Birkhoff-Witt argument that 
$\bar\W$ has a well-defined structure of an associative algebra.


In physics literature the generators of the Weyl algebra $p_i, q_i$ are called
free bosons. The boson-fermion correspondence is a way of
constructing free fermions out of free bosons and vice versa ([9],[14]).
The free fermions are the generators $\psi_i, \psi^*_i$,
 \ $i\in\Z$,
of the Clifford algebra $\Cl$ 
satisfying the relations
$$\psi_i \psi^*_j +  \psi^*_j \psi_i = \delta_{ij}, 
\hbox{\hskip 0.5cm} \psi_i \psi_j +  \psi_j \psi_i = 0,
\hbox{\hskip 0.5cm} \psi^*_i \psi^*_j +  \psi^*_j \psi^*_i = 0, \hbox{\hskip 0.5cm} i,j\in\Z .$$

Form the formal generating series (fermion fields)
$$ \psi(z) = \sum_{k\in\Z} \psi_k z^{-k} \hbox{\rm \ \ and \ \ }
\psi^*(z) = \sum_{k\in\Z} \psi^*_k z^k .$$

{\bf Proposition 1.1.} (cf. [14], Theorem 14.10) The Clifford algebra
$\Cl$ can be represented on the space 
$B = \C[u,u^{-1}]\otimes\bar\W$ by vertex operators
$$\psi(z) \mapsto z^{u{\d \over \d u}} u
\exp\left({1\over 2} \sum_{j\in\N} q_j z^j \right)
\exp\left( {1\over 2} \sum_{j\in\N} p_j z^{-j} \right) ,$$
$$\eqno{(1.5)}$$
$$\psi^*(z) \mapsto u^{-1} z^{- u{\d \over \d u}} 
\exp\left( - {1\over 2} \sum_{j\in\N} q_j z^j \right)
\exp\left(-{1\over 2} \sum_{j\in\N} p_j z^{-j} \right) .$$

Here the correspondence between two formal series is the correspondence
between their respective components.

Note that the components of the vertex operators belong to the algebra 
$B = \C[u,u^{-1}]\otimes\bar\W$ and 
act on $B$ by left multiplication. The symbol $z^{u{\d \over \d u}}$
is interpreted as $z^{u{\d \over \d u}} (u^m) = z^m u^m$.

The only difference between the statement above and Theorem 14.10 in [14] 
is that here we consider the action of these vertex operators 
by left multiplication on $\C[u,u^{-1}]\otimes\bar\W$ and not on
$\C[u,u^{-1}]\otimes F$ as differential operators. Proposition 1.1
is thus an immediate corollary.

The obvious relations between the operator of multiplication by $u$ and
the fields $\psi(z), \psi^*(z)$
$$\psi(z) u = z u \psi(z) , \hbox{\hskip 0.5cm}
\psi^*(z) u = z^{-1} u \psi^*(z) $$
can be rewritten for their components as
$$\psi_k u = u \psi_{k+1} , \hbox{\hskip 0.5cm}
\psi^*_k u = u \psi^*_{k+1} . \eqno{(1.6)} $$

The classical matrix Lie algebras can be embedded into Clifford algebras
and Weyl algebras. It is well-known that this can be also done for
the affine Kac-Moody algebras [8] 
and, in particular, for the affine algebra
$a_\infty$ of infinite rank.

The algebra $a_\infty$ is a non-trivial 
one-dimensional central extension of 
the Lie algebra $\bar a_\infty$ of infinite matrices with finitely many
non-zero diagonals:
$$\bar a_\infty = \left\{ \sum_{i,j\in\Z} a_{ij} E_{ij} 
\hbox{\hskip 0.3cm} \bigg| \hbox{\hskip 0.3cm}
\exists n\in \N \hbox{\hskip 0.5cm} |i-j|>n \imply a_{ij} = 0 
\right\} ,$$
$$a_\infty = \bar a_\infty \oplus \C 1 .$$

The action of $a_\infty$ on $\C[u,u^{-1}]\otimes\bar\W$
is determined by the formula:
$$ E_{ij} \mapsto \left\{ \matrix{
\hbox{\hskip 0.3cm}
\psi_i \psi^*_j \hbox{\rm \ \ if \ \ } i\neq j 
\hbox{\rm \ \ or \ \ }
i = j > 0 ,\cr
- \psi^*_j \psi_i  \hbox{\rm \ \ if \ \ }  
i = j \leq 0 . \hfill \cr}
\right. $$

 Using (1.5) we can write a generating series for this
action ([14], 14.10.9):
$$ \sum\limits_{i,j\in\Z} E_{ij} z_1^i z_2^{-j} \mapsto
{1\over 1 - {z_2\over z_1}} \left(
\left( {z_1\over z_2} \right)^{u{\d\over \d u}}
\Gamma(z_1, z_2) - 1 \right) ,$$
where
$$\Gamma(z_1, z_2) = \exp \left( {1\over 2}
\sum_{j=1}^\infty (z_1^j - z_2^j) q_j \right)
 \exp \left( {1\over 2}
 \sum_{j=1}^\infty (z_1^{-j} - z_2^{-j}) p_j \right) .$$

In order to get a connection with the differential
equations in Hirota form, we should also consider
the tensor square $B\otimes B$ of the $a_\infty$-module
$B = \C[u,u^{-1}]\otimes\bar\W$.

The key link is the Casimir operator ([14], 14.11.2)
$$\Omega = 
\sum\limits_{k\in\Z} \psi_k \otimes \psi_k^*$$
which acts on a completion $\bar{B\otimes B}$ and 
commutes with the $a_\infty$-action ([14], 14.11).
To construct this completion we view $\W\otimes\W$
as a Weyl algebra in twice as many generators and
take its completion as in (1.3), (1.4). The algebra
 $\bar{B\otimes B}$ is the tensor product of 
  $\bar{\W\otimes\W}$ with two copies of $\C[u,u^{-1}]$.

\

\

{\bf 2. Skew Casimir operators}

\

It is well-known that affine Kac-Moody algebra
$\sl = sl_2 \left( \C[t,t^{-1}] \right) \oplus \C K$
may be embedded into $a_\infty$ by
$$ H_{2j+1} \mapsto \sum\limits_{i\in\Z} E_{i,i+2j+1}
\hbox{\hskip 0.2cm} , \hbox{\hskip 0.6cm}
A_j \mapsto \sum\limits_{i\in\Z} (-1)^{i+j} E_{i,i+j}
\hbox{\hskip 0.2cm} , \hbox{\hskip 0.6cm}
K \mapsto 1 
\hbox{\hskip 0.2cm} , \hbox{\hskip 0.6cm}
j \in\Z , \eqno{(2.1)}$$
where
$$ H_{2j+1} = \pmatrix{0 &t^j \cr t^{j+1} &0 \cr}
\hbox{\hskip 0.2cm} , \hbox{\hskip 0.6cm}
A_{2j} = \pmatrix{-t^j  &0 \cr 0 &t^j \cr}
\hbox{\hskip 0.2cm} , \hbox{\hskip 0.6cm}
A_{2j+1} = \pmatrix{0 &t^j \cr -t^{j+1} &0 \cr}
\hbox{\hskip 0.2cm} .$$
The restriction of the representation of $a_\infty$ on $B$
to this subalgebra yields (cf. [14], 14.13):
$$H_k \mapsto -{k\over 2} p_k
\hbox{\hskip 0.2cm} , \hbox{\hskip 0.6cm}
H_{-k} \mapsto {k\over 2} q_k
\hbox{\hskip 0.2cm} , \hbox{\hskip 0.6cm}
k\in\Nodd , \eqno{(2.2)}$$
$$\sum\limits_{j\in\Z} A_j z^{-j} \mapsto
{1\over 2} \left( (-1)^{u{\d \over \d u}} \Gamma (z) - 1
\right) ,\eqno{(2.3)}$$
where
$$ \Gamma (z) = \exp \left(\sum_{j\in\Nodd} q_j z^j
\right) \exp\left(\sum_{j\in\Nodd} p_j z^{-j} \right) 
.\eqno{(2.4)}$$

Lie algebra $\g = \sl$ has a $\Z_2$-grading $\g = \g_0
\oplus \g_1$, where $\g_0$ is the principal Heisenberg
subalgebra spanned by $H_j$ , $j\in\Zodd$ and $K$, while
$\g_1$ is spanned by $A_j + {1\over 2} \delta_{j,0} K$,
$j\in\Z$. The following lemma is an immediate consequence
of the formulas (2.2)-(2.4) above:

{\bf Lemma 2.1.} (i) $x u = u x$ for $x\in \g_0$,

(ii) $x u = - u x$ for $x\in \g_1$.

This result can also be seen from the realization of
elements of $\g$ as infinite matrices. The conjugation
by $u$ is a shift operator: $u^{-1} E_{ij} u =
E_{i+1, j+1} - \delta_{i, 0} \delta_{j,0} 1$.
It follows from (2.1) that
the elements in $\g_0$ are represented by matrices
invariant under this shift, while elements of $\g_1$
are represented by anti-invariant matrices.

The subspace $1\otimes\bar\W$ in $B$ is clearly invariant
under $a_\infty$ and $\sl$-actions. While the Casimir
operator $\Omega$ does not leave the space $\bar\WW$
invariant, it can be easily modified into an operator
$\Lambda_0$, for which $\bar\WW$ is invariant:
$$\Lambda_0 = (1\otimes u) \Omega (u^{-1}\otimes 1) .$$
Moreover using the conjugation by $1\otimes u^k$ 
we obtain a whole family of skew Casimir operators:
$$\Lambda_k = (1\otimes u^{k+1}) \Omega (u^{-1} \otimes 
u^{-k})
\hbox{\hskip 0.2cm} , \hbox{\hskip 0.6cm}
k\in\Z .$$ 
Note that the conjugation by elements $u^k \otimes 1$
produces the same family because $\Omega$ commutes with
$u\otimes u$. 

It is easy to see that
$$\sum_{k\in\Z} \Lambda_k z^{-k} =
\big( \psi(z)u^{-1}\big)\otimes \big( u\psi^*(z) \big).$$

Restricted to $\bar\WW$, the generating series
$\Lambda(z) = \sum\limits_{k\in\Z} \Lambda_k z^{-k}$
can be explicitly written as
$$ \Lambda(z) = \exp \left( {1\over 2} 
\sum_{j\in\N} q_j z^j 
\right) \exp \left( 
{1\over 2} \sum_{j\in\N} p_j z^{-j} \right)
\otimes  \exp \left( - {1\over 2} \sum_{j\in\N} q_j z^j 
\right) \exp \left( - {1\over 2}
\sum_{j\in\N} p_j z^{-j} \right) .$$

The skew Casimir operators $\Lambda_k$ do not anymore 
commute with 
the action of $a_\infty$, but as we see in the following
Proposition, behave nicely relatively to
the subalgebra $\sl$ in $a_\infty$.

{\bf Proposition 2.2.} (a) For $x\in\g_0$
$$( x\otimes 1 + 1\otimes x) \Lambda_k = 
\Lambda_k ( x\otimes 1 + 1\otimes x) , \hbox{\hskip 0.4cm}
k\in\Z.$$
(b) For $x\in\g_1$
$$( x\otimes 1 + 1\otimes x) \Lambda_k = 
- \Lambda_k ( x\otimes 1 + 1\otimes x) , \hbox{\hskip 0.4cm}
k\in\Zodd,$$
$$( x\otimes 1 - 1\otimes x) \Lambda_k = 
- \Lambda_k ( x\otimes 1 - 1\otimes x) , \hbox{\hskip 0.4cm}
k\in\Zev.$$

{\bf Proof.} Recall that $\Lambda_k = (1\otimes u^{k+1}) \Omega
(u^{-1} \otimes u^{-k})$. Part (a) is obvious because 
$x u = u x$ for $x\in\g_0$ by Lemma 2.1 and $x\otimes 1 + 1\otimes x$
commutes with the Casimir operator $\Omega$.

Verification of (b) is also straightforward. Suppose that $x\in\g_1$
and $k$ is odd. By Lemma 2.1, $x$ commutes with the even powers of $u$
and anticommutes with the odd powers of $u$. Thus 
$$( x\otimes 1 + 1\otimes x) (1\otimes u^{k+1}) \Omega 
(u^{-1}\otimes u^{-k})$$ 
$$= (1\otimes u^{k+1}) ( x\otimes 1 + 1\otimes x) \Omega 
(u^{-1}\otimes u^{-k}) $$ 
$$= - (1\otimes u^{k+1}) \Omega (u^{-1}\otimes u^{-k})
( x\otimes 1 + 1\otimes x) .$$ 
If $k$ is even then
$$( x\otimes 1 - 1\otimes x) (1\otimes u^{k+1}) \Omega 
(u^{-1}\otimes u^{-k}) $$ 
$$= (1\otimes u^{k+1}) ( x\otimes 1 + 1\otimes x) \Omega 
(u^{-1}\otimes u^{-k}) $$ 
$$= - (1\otimes u^{k+1}) \Omega (u^{-1}\otimes u^{-k})
( x\otimes 1 - 1\otimes x) .$$

\

Let $\hat\tau = \hat\tau(z_1, z_2,\ldots)$ be a formal Laurent
series in finitely many formal real variables
$z_1, z_2, \ldots$ with coefficients in $\bar\W$.
Let $\hat\tau^*(z_1, z_2,\ldots)$ be the complex conjugate
of $\hat\tau$.

{\bf Proposition 2.3.} The sets of solutions 
$\hat\tau(z_1, z_2,\ldots)$ of equations 
$$ (\hat\tau\otimes\hat\tau) \Lambda_k = \Lambda_k 
(\hat\tau^*\otimes\hat\tau^*) , \hbox{\hskip 0.6cm} k\in\Zodd,
\eqno{(2.5)}$$
and
$$ (\hat\tau\otimes\hat\tau^*) \Lambda_k = \Lambda_k 
(\hat\tau^*\otimes\hat\tau) , \hbox{\hskip 0.6cm} k\in\Zev,
\eqno{(2.6)}$$
are invariant under the transformation
$$\hat\tau(z_1, z_2,\ldots) \mapsto 
\hat\tau^\prime(z_1, z_2,\ldots, z) = 
\hat\tau(z_1, z_2,\ldots) \exp\left(\alpha i \Gamma(z)\right),
$$
where $\alpha\in\R$ and $z$ is a formal real variable.

The equations (2.5) and (2.6) are understood here
as equalities of the corresponding $\bar\WW$-valued
coefficients in the formal Laurent series. 

{\bf Proof.} Since the components of $\Gamma(z)$ 
represent the elements of $\g_1$, we get from
Proposition 2.2 that for $k\in\Zodd$
$$\exp\left(\alpha i \Gamma (z) \right) \otimes
 \exp\left(\alpha i \Gamma (z) \right) \Lambda_k 
=
\exp\left(\alpha i \left( \Gamma(z) \otimes 1
+ 1\otimes \Gamma(z) \right) \right) \Lambda_k $$
$$= \Lambda_k \exp\left( - \alpha i \left( \Gamma(z) 
\otimes 1 + 1\otimes \Gamma(z) \right) \right) =
\Lambda_k \exp\left(- \alpha i \Gamma (z) \right) 
\otimes \exp\left(- \alpha i \Gamma (z) \right) , $$
and in a similar way for $k\in\Zev$:
$$\exp\left(\alpha i \Gamma (z) \right) \otimes
 \exp\left(- \alpha i \Gamma (z) \right) \Lambda_k 
=
\Lambda_k \exp\left(- \alpha i \Gamma (z) \right) 
\otimes \exp\left( \alpha i \Gamma (z) \right). $$

Suppose that $\hat\tau$ is a solution of the equation (2.5)
$$(\hat\tau\otimes\hat\tau) \Lambda_k = \Lambda_k 
(\hat\tau^*\otimes\hat\tau^*) , \hbox{\hskip 0.6cm} k\in\Zodd.
$$
Then for $\hat\tau^\prime = \hat\tau \exp \left( \alpha i 
\Gamma(z) \right)$ we have 
$$ (\hat\tau^\prime\otimes\hat\tau^\prime) \Lambda_k =
 (\hat\tau\otimes\hat\tau) 
\left( \exp\left(\alpha i \Gamma (z) \right) \otimes
\exp\left(\alpha i \Gamma (z) \right) \right) \Lambda_k$$ 
$$ = (\hat\tau\otimes\hat\tau) 
\Lambda_k \left( \exp\left(- \alpha i \Gamma (z) \right) 
\otimes \exp\left(-\alpha i \Gamma (z) \right) \right)$$
$$ = \Lambda_k (\hat\tau^*\otimes\hat\tau^*)
\left( \exp\left(- \alpha i \Gamma (z) \right) 
\otimes \exp\left(-\alpha i \Gamma (z) \right) \right) =
\Lambda_k ((\hat\tau^\prime)^*\otimes(\hat\tau^\prime)^*) .$$
Thus $\hat\tau^\prime$ is also a solution of (2.5).
The case of (2.6) is completely analogous.

{\bf Corollary 2.4.} 
The series $\hat\tau = \left(1 + \alpha_1 i \Gamma(z_1) \right) \ldots
\left(1 + \alpha_N i \Gamma(z_N) \right)$ in formal real variables
$z_1,\ldots, z_N$ with $\alpha_1,\ldots,\alpha_N\in\R$ is a solution
for both (2.5) and (2.6).

{\bf Proof.} Trivially $\hat\tau = 1$ is a solution for both equations.
Thus by the above proposition, 
$\hat\tau = \exp\left(\alpha_1 i \Gamma(z_1) \right) \ldots
\exp\left(\alpha_N i \Gamma(z_N) \right)$ satisfies (2.5) and (2.6).
Finally, $\exp\left(\alpha i \Gamma(z) \right) = 
1 + \alpha i \Gamma(z) $ because $\Gamma^2(z) = 0$
([14], 14.11.15), and the claim of the Corollary follows.

Consider the real and the imaginary parts of $\hat\tau$:
$\hat\tau_0(z_1, z_2, \ldots) = \Re(\hat\tau) ,
 \hat\tau_1(z_1, z_2, \ldots) = \Im(\hat\tau)$.
The equations (2.5) and (2.6) can be rewritten as the systems
$$ \left\{\matrix{
(\hat\tau_0 \otimes \hat\tau_0 - \hat\tau_1 \otimes \hat\tau_1) \Lambda_k -
\Lambda_k (\hat\tau_0 \otimes \hat\tau_0 - \hat\tau_1 \otimes \hat\tau_1) = 0 \cr
(\hat\tau_0 \otimes \hat\tau_1 + \hat\tau_1 \otimes \hat\tau_0) \Lambda_k +
\Lambda_k (\hat\tau_0 \otimes \hat\tau_1 + \hat\tau_1 \otimes \hat\tau_0) = 0 \cr}
\right. \hbox{\hskip 0.5cm} k\in\Zodd \eqno{(2.7)}$$
and
$$ \left\{\matrix{
(\hat\tau_0 \otimes \hat\tau_0 + \hat\tau_1 \otimes \hat\tau_1) \Lambda_k -
\Lambda_k (\hat\tau_0 \otimes \hat\tau_0 + \hat\tau_1 \otimes \hat\tau_1) = 0 \cr
(\hat\tau_0 \otimes \hat\tau_1 - \hat\tau_1 \otimes \hat\tau_0) \Lambda_k +
\Lambda_k (\hat\tau_0 \otimes \hat\tau_1 - \hat\tau_1 \otimes \hat\tau_0) = 0 \cr}
\right. \hbox{\hskip 0.5cm} k\in\Zev . \eqno{(2.8)}$$

The requirement that $\alpha$ and $z$ are real may be dropped 
if we apropriately adjust the statement of Proposition 2.3:

{\bf Proposition 2.5.} The sets of solutions $(\hat\tau_0, \hat\tau_1)$
of the systems (2.7) and (2.8) are invariant under 
the transformation
$$(\hat\tau_0, \hat\tau_1) \mapsto (\hat\tau_0^\prime, \hat\tau_1^\prime) =
(\hat\tau_0, \hat\tau_1) \left(\matrix{
1 &\alpha \Gamma(z) \cr -\alpha \Gamma(z) &1 \cr} \right)$$
where $z$ is a formal variable and $\alpha\in\C$.

The proof parallels the one of Proposition 2.3.

\

\

{\bf 3. Hirota bilinear equations and their solutions}

\

The differential nature of the Weyl algebra is encoded in its defining
relations (1.1). In this section we shall see how to convert purely
algebraic equations
$$ (\hat\tau\otimes\hat\tau) \Lambda_k = \Lambda_k 
(\hat\tau^*\otimes\hat\tau^*) , \hbox{\hskip 0.6cm}
 k\in\Zodd,$$
$$ (\hat\tau\otimes\hat\tau^*) \Lambda_k = \Lambda_k 
(\hat\tau^*\otimes\hat\tau) , \hbox{\hskip 0.6cm} 
k\in\Zev ,$$
into a hierarchy of differential equations in Hirota form
on the ``dequantized'' functions $\tau_0 = \pi(\hat\tau_0), 
\tau_1 = \pi(\hat\tau_1)$.

When a product of two ``functions'' in $\W$ is written in the normally
ordered form, expressions involving partial derivatives (1.2) appear.
One special case of this procedure is recorded in the following Lemma
and will be later used in our calculations.

{\bf Lemma 3.1.} Let $p,q$ be generators of the Weyl algebra
satisfying $pq - qp = c \cdot 1$. Then for $\hat\tau = \hat\tau (q,p)$ we have
$$\exp(zp) \hat\tau = \left\{ \exp\left(cz{\d\over\d q}\right) 
\hat\tau\right\} \exp(zp)$$
and 
$$\hat\tau \exp(zq) = \exp(zq) \left\{ \exp\left(cz{\d\over\d p}\right) 
\hat\tau\right\}. $$
Here $z$ is a formal variable and the above equalities are interpreted
as equalities of formal power series.

{\bf Proof.} We shall prove the second identity, the first will then
follow by applying the automorphism $p \mapsto q, 
q \mapsto -p$.
From (1.2) we get that 
$$ \hat\tau(q,p) q = q \hat\tau(q,p) + c{\d \hat\tau\over \d p} = 
\left( q + c{\d\over\ \d p}\right) \hat\tau(q,p) .$$
By induction we obtain
$$ \hat\tau(q,p) q^n =  \left( q + c{\d\over\ \d p}\right)^n \hat\tau(q,p) .$$
To complete the proof, we apply the above equality to
the Taylor expansion of $\exp(zq)$:
$$ \hat\tau(q,p) \exp(zq) = \exp \left( zq + cz{\d\over\ \d p}\right) 
\hat\tau(q,p) $$
$$ = \exp(zq) \left\{  \exp \left( cz{\d\over\ \d p}\right) 
\hat\tau(q,p) \right\} . $$

At this point we are ready to transform (2.7) and (2.8) 
into a hierarchy of Hirota bilinear equations. 
Note that the terms involved in these equations 
are of the form 
\break
$(f\otimes g) \Lambda_k \pm \Lambda_k (f\otimes g)$.
We shall work with such expressions using the generating series
$(f\otimes g) \Lambda(z) \pm \Lambda(z) (f\otimes g)$, 
$f\otimes g \in\bar\W\otimes\bar\W$.
Denote the variables in the first copy of $\bar\W$ by 
$q^\prime_i, p^\prime_i $ and in the second copy of $\bar\W$ by 
$q^{\prime\prime}_i, p^{\prime\prime}_i$. Then
$$ (f\otimes g) \Lambda(z) \pm \Lambda(z) (f\otimes g) = $$
$$  = f(\q^\prime,\p^\prime) g(\q^{\prime\prime},\p^{\prime\prime})
\exp \left( {1\over 2}
\sum\limits_{j=1}^\infty (q_j^\prime - q_j^{\prime\prime})
z^j \right)
\exp \left( {1\over 2} \sum\limits_{j=1}^\infty 
(p_j^\prime - p_j^{\prime\prime}) z^{-j} \right)$$
$$ \pm 
\exp \left( {1\over 2} 
\sum\limits_{j=1}^\infty (q_j^\prime - q_j^{\prime\prime})
z^j \right)
\exp \left( {1\over 2} \sum\limits_{j=1}^\infty 
(p_j^\prime - p_j^{\prime\prime}) z^{-j} \right)
f(\q^\prime,\p^\prime) g(\q^{\prime\prime},\p^{\prime\prime}). $$

We shall assume that both $f(\q^\prime,\p^\prime)$ and
$g(\q^{\prime\prime},\p^{\prime\prime})$ are in the normal form.
Moreover since $p_i^\prime$ commutes with $q_j^{\prime\prime}$, we
can replace 
$f(\q^\prime,\p^\prime) g(\q^{\prime\prime},\p^{\prime\prime})$
with the normally ordered product \ \ 
$:f(\q^\prime,\p^\prime) g(\q^{\prime\prime},\p^{\prime\prime}):$ \ , \ 
i.e. move all $q$'s to the left of all $p$'s. 

Making the change of variables 
$p_i = {1\over 2} (p_i^\prime + p_i^{\prime\prime}), 
\tp_i = {1\over 2} (p_i^\prime - p_i^{\prime\prime}), 
q_i = {1\over 2} (q_i^\prime + q_i^{\prime\prime}), 
\tq_i = {1\over 2} (q_i^\prime - q_i^{\prime\prime})$,
we rewrite the above expression as
$$ :f(\q + \tbq, \p + \tbp) g(\q - \tbq, \p - \tbp):
\exp \left( \sum\limits_{j=1}^\infty \tq_j z^j \right)
\exp \left( \sum\limits_{j=1}^\infty \tp_j z^{-j} \right)$$
$$\pm \exp \left( \sum\limits_{j=1}^\infty \tq_j z^j \right)
\exp \left( \sum\limits_{j=1}^\infty \tp_j z^{-j} \right)
:f(\q + \tbq, \p + \tbp) g(\q - \tbq, \p - \tbp): .$$

Collecting $\tq$'s on the left and $\tp$'s on the right using 
Lemma 3.1 (note that $[\tp_i, \tq_j] = - {2\over j} \delta_{ij} 
\cdot 1$), we get
\vfill\eject
$$ \exp \left( \sum\limits_{j=1}^\infty \tq_j z^j \right)
\times \hbox{\hskip 10cm}$$
$$\left\{ \left(
\exp 
\left( -2 \sum\limits_{j=1}^\infty {z^j\over j} {\d\over\d\tp_j}\right)
\pm \exp \left( -2 \sum\limits_{j=1}^\infty 
{z^{-j}\over j} {\d\over\d\tq_j}\right) \right)
:f(\q + \tbq, \p + \tbp) g(\q - \tbq, \p - \tbp):
\right\} $$
$$\hbox{\hskip 7cm} \times 
\exp \left( \sum\limits_{j=1}^\infty \tp_j z^{-j} \right) . \eqno{(3.1)}$$

We can convert this expression into the Hirota form. Recall that the 
Hirota bilinear differentiation is defined as
$$ P\left(D_x, D_y, \ldots \right) 
\left[f(x,y,\ldots)\circ g(x,y,\ldots) \right] :=$$
$$P\left({\d\over\d \tx}, {\d\over\d \ty}, \ldots \right)
f(x+\tx, y+\ty,\ldots) g(x - \tx, y - \ty, \ldots) 
\big|_{\tx=0,\ty=0,\ldots} .$$

Using the Taylor formula we get that (see [14], 14.11.8)
$$P\left({\d\over\d \tx}, {\d\over\d \ty}, \ldots \right)
f(x+\tx, y+\ty,\ldots) g(x - \tx, y - \ty, \ldots) =$$
$$P\left(D_x, D_y, \ldots \right) 
\exp\left( \tx D_x + \ty D_y + \ldots \right)
\left[f(x,y,\ldots)\circ g(x,y,\ldots) \right] .$$

Applying this identity we transform (3.1) into
$$ \exp \left( \sum\limits_{j=1}^\infty \tq_j z^j \right)
\left\{ 
\exp \left(\sum\limits_{j=1}^\infty \tq_j D_{q_j} \right)
\left( \exp 
\left( -2 \sum\limits_{j=1}^\infty {z^j\over j} 
D_{p_j}\right)
\pm \exp \left( -2 \sum\limits_{j=1}^\infty 
{z^{-j}\over j} D_{q_j}\right) \right)
\right.$$
$$\exp \left(\sum\limits_{j=1}^\infty \tp_j D_{p_j} \right)
[:f(\q, \p) \circ g(\q , \p):]
\Bigg\}
\exp \left( \sum\limits_{j=1}^\infty \tp_j z^{-j} \right) . \eqno{(3.2)}$$

We use the reduction to the subalgebra $\sl$ in 
$a_\infty$ to construct solutions of (2.5) and (2.6). 
We see from (2.2)-(2.4) that this reduction
involves only variables $p_j$, $q_j$ with $j\in\Nodd$, so
the solutions we get in Corollary 2.4 do not depend on
the even-indexed variables. Thus in (3.2) we may set
$D_{p_j} = 0$, $D_{q_j} = 0$ for $j\in\Nev$.
In (3.2) all $q$'s are collected to the left of all $p$'s
and all $\tq$'s are to the left of all $\tp$'s.
Because of that we can easily evaluate the image of this
expression under the map $\pi$, which gives us:
$$ R_{\pm}(z) [\pi(f)(\x) \circ \pi(g)(\x)] , $$
where 
$$ R_{\pm}(z) = 
\exp \left( \sum\limits_{j\in \Z \backslash \{ 0 \} }
 \tx_j z^j \right)
\exp \left(\sum\limits_{j\in\Zodd} \tx_j D_{x_j} \right)
\hbox{\hskip 3cm}$$
$$ \hbox{\hskip 3cm}
\times \left\{
\exp \left( -2 \sum\limits_{j\in\Nodd} {z^j\over j} 
D_{x_{-j}}\right)
\pm \exp \left( -2 \sum\limits_{j\in\Nodd} 
{z^{-j}\over j} D_{x_j}\right) \right\}
$$
and $\x = ( \ldots, x_{-3}, x_{-1}, x_1, x_3 , \dots)$.
Here
$R_\pm(z)$ is a Laurent series in $z$ and a Taylor series 
in  $\tx_j$. The coefficients of this series are 
Hirota differential operators. 

Now we can rewrite (2.7) and (2.8)
as the generating series for the hierarchy of 
Hirota bilinear equations on $\tau_0 = \pi(\hat\tau_0)$
and $\tau_1 = \pi(\hat\tau_1)$:

$$\Res \left( z^j R_-(z) \right)
(\tau_0 \circ \tau_0 - 
\tau_1 \circ \tau_1 ) = 0 , \hbox{\hskip 0.6cm}
j\in\Zev ,\eqno{(3.3)}$$

$$\Res \left( z^j R_+(z) \right)
(\tau_0 \circ \tau_1 + 
\tau_1 \circ \tau_0 ) = 0 , \hbox{\hskip 0.6cm}
j\in\Zev ,\eqno{(3.4)}$$

$$\Res \left( z^j R_-(z) \right)
(\tau_0 \circ \tau_0 + 
\tau_1 \circ \tau_1 ) = 0 , \hbox{\hskip 0.6cm}
j\in\Zodd ,\eqno{(3.5)}$$

$$\Res \left( z^j R_+(z) \right)
(\tau_0 \circ \tau_1 - 
\tau_1 \circ \tau_0 ) = 0 , \hbox{\hskip 0.6cm}
j\in\Zodd . \eqno{(3.6)}$$
As usual, the residue denotes the coefficient at $z^{-1}$
of a formal Laurent series. The coefficient at each
monomial in $\tx$'s in the formal equations above
is a Hirota bilinear equation on $\tau_0$
and $\tau_1$. We shall study the Hirota equations that 
occur here in the next section. Now we turn to describing
solutions of this hierarchy.

From Corollary 2.4 it follows that
$$\tau_0 + i\tau_1 =
\pi \left(  \left( 1 + \alpha_1 i \Gamma(z_1) \right)
\ldots  \left( 1 + \alpha_N i \Gamma(z_N) \right) 
\right)$$
is a solution of (3.3) - (3.6). From [14] we get that the 
right hand side converges to 
$$ 1 + \sum_{k=1}^N \sum_{1\leq j_1 < \ldots < j_k\leq N}
\alpha_{j_1} \ldots \alpha_{j_k} i^k 
\prod\limits_{1\leq r < s \leq k}
\left( {{z_{j_r} - z_{j_s}} \over {z_{j_r} + z_{j_s}}} \right)^2
\exp\left( \sum_{m\in\Zodd} \left(\sum_{s=1}^k z_{j_s}^m \right) x_m
\right) \eqno{(3.7)} $$
when $|z_1| > |z_2| > \ldots > |z_N|$. By analytic continuation
we conclude that
\vfill\eject
$$ \tau_0 = 1 + \sum_{k=1}^{[{N \over 2}]} (-1)^k
 \sum_{1\leq j_1 < \ldots < j_{2k}\leq N}
\alpha_{j_1} \ldots \alpha_{j_{2k}} 
\prod\limits_{1\leq r < s \leq 2k}
\left( {{z_{j_r} - z_{j_s}} \over {z_{j_r} + z_{j_s}}} \right)^2$$
$$\times
\exp\left( \sum_{m\in\Zodd} \left(\sum_{s=1}^{2k}
 z_{j_s}^m \right) x_m
\right) , \eqno{(3.8)} $$
$$ \tau_1 =  \sum_{k=0}^{[{N-1 \over 2}]} (-1)^k
\sum_{1\leq j_1 < \ldots < j_{2k+1}\leq N}
\alpha_{j_1} \ldots \alpha_{j_{2k+1}}  
\prod\limits_{1\leq r < s \leq {2k+1}}
\left( {{z_{j_r} - z_{j_s}} \over {z_{j_r} + z_{j_s}}} \right)^2$$ 
$$\times
\exp\left( \sum_{m\in\Zodd} \left(\sum_{s=1}^{2k+1}
 z_{j_s}^m \right) x_m
\right) \eqno{(3.9)} $$
is a solution of the hierarchy (3.3)-(3.6) for all $z_1, \dots z_N$.

As we shall see in the next section, this hierarchy contains the
sine-Gordon equation, as well as two copies of the Korteweg - de Vries
hierarchy.

\

\

{\bf 4. The KdV -- sine-Gordon -- KdV hierarchy}

\

In this section we transform the Hirota equations (3.3)-(3.6)
into non-linear PDEs. We show that among other equations,
this hierarchy contains the sine-Gordon equation, the KdV and 
the modified KdV equations. We begin by listing some of the Hirota
equations that are coefficients at monomials in (3.3)-(3.6) (we 
multiply them by appropriate constants to avoid fractional
coefficients). For simplicity of notations, we write $D_j$ for
$D_{x_j}$.

 From the series (3.3):
$$\eqalignno{
 D_{-1} D_1 
(\tau_0 \circ \tau_0 - \tau_1 \circ \tau_1 ) &= 0
\hbox{{\hskip 1cm} ({\rm at }} x_1 z), 
&(4.1) \cr
\left( D_1^4 - D_1 D_3 \right)
(\tau_0 \circ \tau_0 - \tau_1 \circ \tau_1 ) &= 0
\hbox{{\hskip 1cm} ({\rm at }} x_1 z^{-3}), 
&(4.2) \cr
\left( D_{-1} D_3 + 2 D_{-1} D_1^3 - 3 D_1^2 \right)
(\tau_0 \circ \tau_0 - \tau_1 \circ \tau_1 ) &= 0
\hbox{{\hskip 1cm} ({\rm at }} x_{-1} z^{-3}), 
&(4.3) \cr
\left( D_{-1} D_3 + D_1^2 \right)
(\tau_0 \circ \tau_0 - \tau_1 \circ \tau_1 ) &= 0
\hbox{{\hskip 1cm} ({\rm at }} x_3 z). 
&(4.4) \cr
%
\noalign{\hbox{ From the series (3.4):}}
\left( D_{-1} D_1 - 1 \right)
(\tau_1 \circ \tau_0) &= 0
\hbox{{\hskip 1cm} ({\rm at }} x_1 z), 
&{(4.5)} \cr
 \left( D_1^4  - D_1 D_3 \right)
(\tau_1 \circ \tau_0) &= 0
\hbox{{\hskip 1cm} ({\rm at }} x_1 z^{-3}), 
&{(4.6)} \cr
 \left( D_1^2  - D_{-1} D_3 \right)
(\tau_1 \circ \tau_0) &= 0
\hbox{{\hskip 1cm} ({\rm at }} x_3 z), 
&{(4.7)} \cr
 \left( D_1^2  - D_{-1} D_1^3 \right)
(\tau_1 \circ \tau_0) &= 0
\hbox{{\hskip 1cm} ({\rm at }} x_1^3 z). 
&{(4.8)} \cr
\noalign{\hbox{ From the series (3.5):}}
D_1^2 
(\tau_0 \circ \tau_0 + \tau_1 \circ \tau_1 ) &= 0
\hbox{{\hskip 1cm} ({\rm at }} x_1^2 z^0), 
&{(4.9)} \cr
 \left( D_1^4 + 2 D_1 D_3 \right)
(\tau_0 \circ \tau_0 + \tau_1 \circ \tau_1 ) &= 0
\hbox{{\hskip 1cm} ({\rm at }} x_1 x_3 z^0), 
&{(4.10)} \cr
 \left( D_{-1} D_1^3 - D_{-1} D_3 \right)
(\tau_0 \circ \tau_0 + \tau_1 \circ \tau_1 ) &= 0
\hbox{{\hskip 1cm} ({\rm at }} x_{-1} x_1 z^{-2}). &{(4.11)} \cr
\noalign{\hbox{ 
And finally from the series (3.6):}}
\left( D_1^3  - D_3 \right)
(\tau_1 \circ \tau_0) &= 0
\hbox{{\hskip 1cm} ({\rm at }} x_1 z^{-2}), 
&{(4.12)}\cr
 \left( D_{-1} D_1^2  - D_1 \right)
(\tau_1 \circ \tau_0) &= 0
\hbox{{\hskip 1cm} ({\rm at }} x_{-1} z^{-2}), 
&{(4.13)} \cr
 \left( D_1^5  - D_5 \right)
(\tau_1 \circ \tau_0) &= 0
\hbox{{\hskip 1cm} ({\rm at }} x_1 z^{-4}), 
&{(4.14)}\cr
 \left( 2 D_1^5  - 5 D_1^2 D_3 + 3 D_5 \right)
(\tau_1 \circ \tau_0) &= 0
\hbox{{\hskip 1cm} ({\rm at }} x_3 z^{-2}). 
&{(4.15)} \cr}$$

One of the transformations that we consider below is
$u = 4 \arctan \left( {\tau_1 \over \tau_0} \right)$
or, equivalently, 
${\tau_1 \over \tau_0} = \tan \left( {u \over 4} \right)$.
Let more generally ${\tau_1 \over \tau_0} = \varphi (u)$,
hence $\tau_1 = \varphi \tau_0$. 

In order to rewrite systems of Hirota equations as 
non-linear PDEs
in function $u$, we need the Leibnitz rule Lemma for the
Hirota differential operators. For a multi-index 
$\beta = (\beta_1, \ldots , \beta_n)$ let
$D^\beta = D_{x_1}^{\beta_1} \ldots  D_{x_n}^{\beta_n} $ be
the Hirota differential operator and let
$\d^\beta = \d_{x_1}^{\beta_1} \ldots  \d_{x_n}^{\beta_n} $ be
the usual differential operator. For a pair of multi-indices
$\beta$ and $\gamma$ with $0\leq \gamma_j \leq \beta_j$ let
$\pmatrix{\beta \cr \gamma \cr} =
\pmatrix{\beta_1 \cr \gamma_1 \cr} \ldots
\pmatrix{\beta_n \cr \gamma_n \cr} .$

{\bf Lemma 4.1.}

(a) $D^\beta (\varphi f \circ g) = \sum\limits_{\gamma}
\pmatrix{\beta \cr \gamma \cr} \d^\gamma (\varphi)
D^{\beta-\gamma} (f \circ g) .$

(b) $D^\beta (\varphi_1 f \circ \varphi_2 g) = \sum\limits_{\gamma}
\pmatrix{\beta \cr \gamma \cr} D^\gamma (\varphi_1 \circ \varphi_2)
D^{\beta-\gamma} (f \circ g) .$

The proof of this Lemma is straightforward.

{\it Sine-Gordon equation.} Consider (4.1) and (4.5):
$$\left\{ \eqalign{
D_{-1} D_1 
(\tau_0 \circ \tau_0 - \tau_1 \circ \tau_1 ) &= 0 \cr
\left( D_{-1} D_1 - 1 \right)
(\tau_1 \circ \tau_0) &= 0 \cr}
\right. \eqno{(4.16)}$$
and set $\tau_1 = \varphi \tau_0$. Applying Lemma 4.1, we
transform this system into
$$\left\{ \eqalign{
(1-\varphi^2) D_{-1} D_1 (\tau_0 \circ \tau_0)
-  D_{-1} D_1 (\varphi \circ \varphi) \tau_0^2 &= 0 \cr
\varphi D_{-1} D_1 (\tau_0 \circ \tau_0)
+ (\d_{-1} \d_1 (\varphi) - \varphi) \tau_0^2 &= 0 .\cr }
\right. $$
Interpreting this as a system of linear equations on
$D_{-1} D_1 (\tau_0 \circ \tau_0)$ and $ \tau_0^2$ we
get that
$$ \det \pmatrix{
1 - \varphi^2 & - D_{-1}D_1 (\varphi \circ \varphi) \cr
\varphi & \d_{-1} \d_1 (\varphi) - \varphi \cr} = 0
\eqno{(4.17)} $$
since this system has a non-trivial solution.

If we now substitute $\varphi = \tan \left( {u\over 4} \right)$
and multiply (4.17) by $\cos^4 \left( {u\over 4} \right)$
then we obtain the sine-Gordon equation:
$$ {\d^2 u \over \d x_{-1} \d x_1} = \sin (u).
\eqno{(4.18)} $$

If we let parameters $\alpha_j$ in (3.8), (3.9) be purely
imaginary then $\tau_1$ will become purely imaginary, while
$\tau_0$ will remain real. In this case $u$ will be purely
imaginary, $u = i\u$, ${\tau_1 \over i\tau_0}
= \tanh \left( {\u \over 4 }\right)$, and the function
$\u$ satisfies the sinh-Gordon equation:
$$ {\d^2 \u \over \d x_{-1} \d x_1} = \sinh (\u).
\eqno{(4.19)} $$

{\it Modified KdV equation.} The same technique applied to
(4.9) and (4.12) 
$$ \left\{ \eqalign{
D_1^2  (\tau_0 \circ \tau_0 + \tau_1 \circ \tau_1 ) &= 0 \cr
\left( D_1^3  - D_3 \right) (\tau_1 \circ \tau_0) &= 0  \cr}
\right. \eqno{(4.20)} $$
yields
$$ \left\{ \eqalign{
(1 + \varphi^2) D_1^2 (\tau_0 \circ \tau_0)
 + D_1^2 (\varphi \circ \varphi) \tau_0^2 &= 0 \cr
 3 \d_1(\varphi) D_1^2 (\tau_0 \circ \tau_0)
 + (\d_1^3(\varphi) - \d_3(\varphi) )
 (\varphi \circ \varphi) \tau_0^2 &= 0 . \cr} 
\right.
\eqno{(4.21)}$$
After the same substitution as above, $\varphi = \tan 
\left( u \over 4 \right)$, multiplying the determinant
of this system by $\cos^4 \left( {u\over 4} \right)$,
we get the mKdV equation:
$$ {\d u \over \d x_3} = {1\over 2} \left({\d u \over \d x_1}\right)^3
+ {\d^3 u \over \d x_1^3}  , 
\eqno{(4.22)} $$
or for $v = {\d u \over \d x_1}$:
$$ {\d v \over \d x_3} = {3\over 2} v^2 {\d v \over \d x_1}
+ {\d^3 v \over \d x_1^3}.
\eqno{(4.23)} $$

{\it Mixed sine-Gordon -- mKdV equation.}
The system of three equations (4.4), (4.7) and (4.9)
$$ \left\{ \eqalign{
\left( D_1^2 + D_{-1} D_3  \right)
(\tau_0 \circ \tau_0 - \tau_1 \circ \tau_1 ) &= 0 \cr
\left( D_1^2  - D_{-1} D_3 \right)
(\tau_1 \circ \tau_0) &= 0 \cr
D_1^2 (\tau_0 \circ \tau_0 + \tau_1 \circ \tau_1 ) &= 0 \cr }
\right. \eqno{(4.24)} $$
becomes a system of three linear equations on
$D_1^2 (\tau_0 \circ \tau_0)$ , \ 
$D_{-1} D_3 (\tau_0 \circ \tau_0)$, \ $\tau_0^2$ 
and from its determinant
$$ \det \pmatrix{
1 - \varphi^2 & 1 - \varphi^2 
& (D_1^2 + D_{-1} D_3) (\varphi \circ \varphi) \cr
\varphi & - \varphi & \d_1^2(\varphi) - \d_{-1} \d_3 (\varphi) \cr
1 + \varphi^2 & 0 & D_1^2 (\varphi \circ \varphi) \cr }
= 0 \eqno{(4.25)}$$
after substitution $\varphi = \tan \left( u \over 4 \right)$ we
obtain the equation
$$ {\d^2 u \over \d x_{-1} \d x_3} = 
{1\over 2} \sin(u) \left({\d u \over \d x_1}\right)^2
+ \cos(u) {\d^2 u \over \d x_1^2}  . 
\eqno{(4.26)} $$
One can  show that this equation follows from the 
sine-Gordon and mKdV equations (4.18), (4.22) 
taken together, however (4.26) may have an
independent physical meaning.

{\it The second mKdV equation.} The system of Hirota equations
(4.9), (4.10), (4.14) and (4.15)
$$ \left\{ \eqalign{
\left( D_1^5  - D_5 \right) (\tau_1 \circ \tau_0) &= 0 \cr
\left( 2 D_1^5  - 5 D_1^2 D_3 + 3 D_5 \right)
(\tau_1 \circ \tau_0) &= 0 \cr
D_1^2 (\tau_0 \circ \tau_0 + \tau_1 \circ \tau_1 ) &= 0 \cr
\left( D_1^4 + 2 D_1 D_3 \right)
(\tau_0 \circ \tau_0 + \tau_1 \circ \tau_1 ) &= 0 \cr}
\right. \eqno{(4.27)} $$
applying the same method as above and eliminating $x_3$ using (4.23),
is transformed into
the following equation for $v = 4 {\d \over \d x_1} 
\arctan \left( {\tau_1 \over \tau_0} \right)$:
$$ {\d v \over \d x_5} = {\d \over \d x_1} \left(
{3\over 8} v^5
+ {5\over 2} v^2 {\d^2 v \over \d x_1^2}
+ {5\over 2} v  \left( {\d  v \over \d x_1}\right)^2
+ {\d^4 v \over \d x_1^4} \right) , 
\eqno{(4.28)} $$
which is essentially the second equation in the mKdV hierarchy
constructed by the AKNS method [1].

{\it The double KdV hierarchy.} It can be easily seen that the
hierarchy (3.3)-(3.6) is invariant under the transformation
$z \mapsto z^{-1}, x_j \mapsto x_{-j}$, i.e., it is symmetric
in the positive and negative directions.

If in (3.3) and (3.4) we set $\tilde x_k = 0$ for $k < 0$ 
and consider non-negative values of $j$,
then
$$\Res \left( z^j R_\pm (z) \right) = 
\pm \Res \left( z^j P(z) \right) , $$
where 
$$P(z) = 
\exp \left( \sum\limits_{j = 1 }^\infty \tx_j z^j \right)
\exp \left(\sum\limits_{j\in\Nodd} \tx_j D_{x_j} \right)
\exp \left( -2 \sum\limits_{j\in\Nodd} {z^{-j}\over j} 
D_{x_j}\right) .
\eqno{(4.29)}$$
 In this case the Hirota polynomials appearing in (3.3) and (3.4)
differ only by sign. Combining (3.3) and (3.4) together we get
that $\tau = \tau_0 + i \tau_1$ satisfies
$$  \Res \left( z^j P(z) \right) (\tau \circ \tau) = 0
\hbox{{\hskip 1cm} {\rm for \ \ }}
j \in \Nev \cup \{ 0 \} . \eqno{(4.30)} $$ 
The Hirota equation 
$$ (D_1^4 - D_1 D_3) (\tau \circ \tau) = 0 \eqno{(4.31)}$$
(cf. (4.2), (4.6)) after the substitution
$ f = {\d^2 \over \d x_1^2} \ln (\tau)$
becomes the Korteweg - de Vries equation
$$ {\d f \over \d x_3} = 12 f {\d f \over \d x_1} 
+ {\d^3 f \over \d x_1^3} . \eqno{(4.32)} $$
The series (4.30) is the KdV hierarchy [4], [14] 
in variables $x_1, x_3, x_5 \ldots$. 
By symmetry, the second copy of the KdV hierarchy extends in the 
negative direction $x_{-1}, x_{-3}, \ldots$ . 
These two KdV hierarchies are linked together by the sine-Gordon
equation (4.18) which involves variables $x_1$ and $x_{-1}$.
The soliton
solutions for the double KdV hierarchy are given by (3.7) 
with purely imaginary values of $\alpha_1, \ldots \alpha_N$.

{\bf Remark 4.2.} Another way to construct this subhierarchy
is to consider the Casimir operator $\Omega_{\rm aff}$ for affine Kac-Moody
algebra $\sl$ (see [14]) acting on $\bar\WW$. The double KdV hierarchy
then arises from the equation
$$(\hat\tau \otimes \hat\tau) \Omega_{\rm aff} = \Omega_{\rm aff} (\hat\tau \otimes \hat\tau) .$$

The double KdV hierarchy that we obtain here contains more than
just two copies of the KdV hierarchy. It also contains equations
that involve both positive and negative variables. The simplest such
an equation on the function $\tau = \tau_0 + i\tau_1$ is
$$ \left( D_{-1} D_3 - 4 D_{-1} D_1^3 + 3 D_1^2 \right)
( \tau \circ \tau ) = 0  ,
\eqno{(4.33)} $$
which follows from (4.3), (4.4), (4.7) and (4.8). 
A similar equation (without the last term) appears
in the $D_4^{(1)}$-hierarchy [15], but the solutions 
constructed here and in [15] are inequivalent.

After the substitution $g = {\d \over \d x_1} \ln (\tau)$ we get
$$ {\d^2 g \over \d x_{-1} \d x_3} =
{\d \over \d x_1} \left( 4 {\d^3 g \over \d x_{-1} \d x_1^2}
+ 24 {\d g \over \d x_{-1} } {\d g \over \d x_1 } 
 - 3 {\d g \over \d x_1 } \right) , \eqno{(4.34)}$$
which is a generalization of the KdV with two spatial variables,
but different from the Kadomtsev-Petviashvili equation.

It is interesting to note that the same equation appears in our
hierarchy in a different way. If we combine (4.3), (4.4) with
(4.9) and (4.11) then we get
$$\left( D_{-1} D_3 - D_{-1} D_1^3 + 3 D_1^2 \right)
( \tau_0 \circ \tau_0 ) = 0  
\eqno{(4.35)} $$
and
$$\left( D_{-1} D_3 - D_{-1} D_1^3 + 3 D_1^2 \right)
( \tau_1 \circ \tau_1 ) = 0 .  
\eqno{(4.36)} $$
These can be transformed into (4.33) by rescaling the variables
$x_{j} = {1\over 2} x_{j}^\prime .$ Thus in addition to the solutions
$g = {\d \over \d x_1} \ln (\tau)$ of (4.34) with $\tau$ given
by (3.7), we also get solutions
$$ \tau  = 1 + \sum_{k=1}^{[{N \over 2}]} (-1)^k
 \sum_{1\leq j_1 < \ldots < j_{2k}\leq N}
\alpha_{j_1} \ldots \alpha_{j_{2k}} 
\prod\limits_{1\leq r < s \leq 2k}
\left( {{z_{j_r} - z_{j_s}} \over {z_{j_r} + z_{j_s}}} \right)^2$$
$$ \times
\exp\left( 
{1 \over 2} \left(\sum_{s=1}^{2k} z_{j_s}^{-1} \right)
 x_{-1} +
{1 \over 2} \left(\sum_{s=1}^{2k} z_{j_s} \right) x_1 +
{1 \over 2} \left(\sum_{s=1}^{2k} z_{j_s}^3 \right) x_3 
\right)  \eqno{(4.37)} $$
and
$$ \tau  =  \sum_{k=0}^{[{N-1 \over 2}]} (-1)^k
\sum_{1\leq j_1 < \ldots < j_{2k+1}\leq N}
\alpha_{j_1} \ldots \alpha_{j_{2k+1}}  
\prod\limits_{1\leq r < s \leq {2k+1}}
\left( {{z_{j_r} - z_{j_s}} \over {z_{j_r} + z_{j_s}}} \right)^2 $$
$$ \times
\exp\left( 
{1 \over 2} \left(\sum_{s=1}^{2k+1} z_{j_s}^{-1} \right) x_{-1} +
{1 \over 2} \left(\sum_{s=1}^{2k+1} z_{j_s} \right) x_1 +
{1 \over 2} \left(\sum_{s=1}^{2k+1} z_{j_s}^3 \right) x_3 
\right) . \eqno{(4.38)} $$

{\it Miura transformation.}
 It is well-known that solutions of the mKdV equation
$$ {\d \v \over \d x_3} =
- {3 \over 2} \v^2 {\d \v \over \d x_1} 
+ {\d^3 \v \over \d x_1^3}  \eqno{(4.39)}$$
with $\v = {1\over i} v$ (cf. (4.23) )
can be transformed into solutions of the KdV equation
$$ {\d f \over \d x_3} = 12 f {\d f \over \d x_1} 
+ {\d^3 f \over \d x_1^3}  \eqno{(4.40)} $$
by the Miura substitution
$$ f = {1 \over 4} {\d \v \over \d x_1}
- {1 \over 8} \v^2  . \eqno{(4.41)} $$
 The soliton solutions of (4.39) and (4.40)
are given by 
$ \v = 4 {\d \over \d x_1} \arctanh 
\left( {1 \over i } {\tau_1 \over \tau_0} \right)$ 
\break
$= 2 {\d \over \d x_1} \ln
\left( {\tau_0 + i \tau_1 \over  \tau_0 - i \tau_1}
\right)$
and 
$ f = {\d^2 \over \d x_1^2} \ln (\tau_0 + i \tau_1)$,
where $\tau_0$ and $\tau_1$ are given by (3.8) and (3.9)
with purely imaginary parameters $\alpha_j$. 
Note that $\tau_0 \pm i \tau_1$ is real in this case.

A natural question to ask is whether for a given pair
$(\tau_0, \tau_1)$, the functions $\v$ and $f$ are
linked by the Miura transform (4.41). The answer to this
question is positive, and as we see from the following Lemma,
the Miura transform is a feature of the whole hierarchy
and not specific just to the pair of the KdV and mKdV
equations.

{\bf Lemma 4.3.}  The functions 
$ \v =  2 {\d \over \d x_1} \ln
\left( {\tau_0 + i \tau_1 \over  \tau_0 - i \tau_1}
\right)$ 
and
$ f = {\d^2 \over \d x_1^2} \ln (\tau_0 + i \tau_1)$
are related by the Miura transform (4.41) if and
only if the Hirota equation (4.10) holds:
$$ D_1^2 (\tau_0 \circ \tau_0 + \tau_1 \circ \tau_1) = 0 .$$ 
The proof of this Lemma is straightforward.

\

\

{\bf References}

\

\noindent
[1]. Ablowitz, M.J., Clarkson, P.A.: Solitons, nonlinear
evolution equations and inverse scattering.
London Mathematical Society Lecture Notes Series 
{\bf 149}.
Cambridge: Cambridge University Press 1991.

\noindent
[2]. Date, E., Jimbo, M., Kashiwara, M., Miwa, T.:
Operator approach to the Kadomtsev-Petviashvili equation.
Transformation groups for soliton equations III.
J. Phys. Soc. Japan {\bf 50}, 3806-3812 (1981).

\noindent
[3]. Date, E., Jimbo, M., Kashiwara, M., Miwa, T.:
Transformation groups for soliton equations IV.
A new hierarchy of soliton equations of KP-type.
Physica 4D 343-365 (1982).

\noindent
[4]. Date, E., Jimbo, M., Kashiwara, M., Miwa, T.:
Transformation groups for soliton equations.
Euclidean Lie algebras and reduction of the KP hierarchy.
Publ. RIMS, Kyoto Univ. {\bf 18} 1077-1110 (1982).

\noindent
[5]. Dixmier, J., Alg\`ebres envellopantes. 
Paris: Gauthier-Villars, 1974.

\noindent
[6]. Drinfeld, V.G., Sokolov, V.V.: Equations of KdV type
and simple Lie algebras, 
Sov. Math. Dokl. {\bf 23} 457-462 (1981).

\noindent
[7]. Enriquez, B., Frenkel, E.: Equivalence of two
approaches to integrable hierarchies of KdV type.
Comm. Math. Phys. {\bf 185} 211-230 (1997).

\noindent
[8]. Feingold, A.J., Frenkel, I.B.: Classical affine
algebras. Adv. Math. {\bf 56}, 117-172 (1985).

\noindent
[9]. Frenkel, I.: Two constructions of affine Lie algebra
representations and boson-fermion correspondence in
quantum field theory.
J. Func. Anal. {\bf 44} 259-327 (1981).

\noindent
[10]. Frenkel, I.B., Kac, V.G.: Basic representations of 
affine Lie algebras and dual resonance models. Invent. 
Math. {\bf 62}, 23-66 (1980).

\noindent
[11]. Hirota, R.: Exact solution of the 
Korteweg - de~Vries
equation for multiple collisions of solitons.
Phys. Rev. Lett. {\bf 27}, 1192-1194 (1971).

\noindent
[12]. Hirota, R.: Exact solution of the modified
Korteweg - de~Vries
equation for multiple collisions of solitons.
J. Phys. Soc. Japan {\bf 33}, 1456-1458 (1972).

\noindent
[13]. Hirota, R.: Exact solution of the sine-Gordon
equation for multiple collisions of solitons.
J. Phys. Soc. Japan {\bf 33}, 1459-1463 (1972).

\noindent
[14]. Kac, V.G.: Infinite dimensional Lie algebras.
3rd ed. Cambridge: Cambridge University Press 1990.

\noindent
[15]. Kac, V.G., Wakimoto, M.: Exceptional hierarchies
of soliton equations. 
Proc. Symp. in Pure Math. {\bf 49} 191-237 (1989).

\noindent
[16]. Khesin, B.A., Zakharevich, I.S.: Poisson-Lie group
of pseudodifferential symbols.
Comm. Math. Phys. {\bf 171} 475-530 (1995).

\noindent
[17]. Lepowsky, J., Wilson, R.L.: Construction of affine 
Lie algebra $A_1^{(1)}$.
Comm. Math. Phys. {\bf 62} 43-53 (1978).

\noindent
[18]. Mandelstam, S.: Soliton operators for the quantized
sine-Gordon equation. Phys. Rev. D. {\bf 11} 3026-3030
(1975).

\noindent
[19]. Sato, M.: The KP hierarchy and infinite-dimensional
Grassmann manifolds. 
Proc. Symp. in Pure Math. {\bf 49} 51-66 (1989).

\end